\documentclass[prl,twocolumn]{revtex4}
\usepackage{amsfonts}
\usepackage{amsmath}
\usepackage{amssymb}
\usepackage{graphicx}

\input{tcilatex}
\begin{document}

\preprint{}
\title[Te eMChA]{Strong electrical magneto-chiral anisotropy in tellurium}
\author{G.L.J.A. Rikken\medskip }
\affiliation{Laboratoire National des Champs Magn\'{e}tiques Intenses\\
UPR3228 CNRS/EMFL/INSA/UGA/UPS, Toulouse \& Grenoble, France.}
\author{N. Avarvari}
\affiliation{MOLTECH-Anjou, UMR 6200, CNRS, UNIV Angers, 2 blvd Lavoisier, 49045 ANGERS
Cedex, France}

\begin{abstract}
We report the experimental observation of strong electrical magneto-chiral
anistropy (eMChA) in trigonal tellurium (t-Te) crystals. We introduce the
tensorial character of the effect and determine several tensor elements and
we \ propose a novel intrinsic bandstructure-based mechanism for eMChA which
gives a reasonable description of the principal results.
\end{abstract}

\date{\today }
\startpage{101}
\endpage{102}
\maketitle

Chirality is important in many areas of physics, chemistry and biology,
where objects or materials can exist in two non-superimposable forms, one
being the mirror image of the other (enantiomers). Such a system is not
invariant under parity reversal. If in addition it is not invariant under
time-reversal because it has a magnetization, an entire class of effects
called magneto-chiral anisotropy (MChA) becomes allowed. The existence of
MChA in the optical properties of chiral systems under magnetic field has
been predicted since 1962 \cite{groenewege},\cite{burstein},\cite{baranova},%
\cite{wagniere},\cite{barron}. It corresponds to a difference in the
absorption and refraction of unpolarized light propagating parallel or
anti-parallel to the magnetic field in a chiral medium. After its first
experimental observations \cite{Naturemca},\cite{kleindienst},\cite{mcaabs},
its existence was confirmed across the entire electromagnetic spectrum, from
microwaves \cite{Microwave MChA} to X-rays \cite{Xray MChA}. The existence
of MChA was further generalized to other transport phenomena \cite{emchaprl}%
. It was experimentally observed in the electrical transport in bismuth
helices \cite{emchaprl}, carbon nano tubes \cite{eMChA CNT}, \cite{Cobden}
and bulk organic conductors \cite{Pop}, as an electrical resistance $R$ that
depends on the handedness of the conductor and on the relative orientation
of electrical current $\mathbf{I}$ and magnetic field\textbf{\ }$\mathbf{B}$%
: \ $R^{D/L}(\mathbf{B},\mathbf{I})=R_{0}(1+\mu ^{2}B^{2}+\gamma ^{D/L}%
\mathbf{B\cdot I)}$, with $\gamma ^{D}=-\gamma ^{L}$ referring to the right-
and left-handed enantiomer of the conductor. More recently, such electrical
MChA  was also observed in chiral magnetic systems \cite{Yokouchi},\cite%
{Maurenbrecher},\cite{Aoki}, which was tentatively explained by scattering
by chiral spin fluctuations. The spin filter effect \cite{Naaman Rev} can
also be regarded as a form of eMChA. As another generalization, MChA was
recently observed in the propagation of ultrasound \cite{US MChA}. Although
the characteristics of eMChA make it interesting for applications, the small
values reported so far daunt their development. Whereas a microscopic theory
exists for optical MChA \cite{barron}, apart from some simplified model
calculations \cite{Free electron helix}, no quantitative theory exists for
eMChA in bulk materials. Here we report the experimental observation of much
larger values in  t-Te, and introduce a novel intrinsic mechanism for eMChA
that explains this observation and allows to identify other materials
showing a strong effect.\newline
Chiral t-Te is an ideal material to study eMChA. It is a well-characterized
direct gap semiconductor (E$_{g}$ = 0,33 eV), with its conduction band
minimum and valence band maximum at the H-point in the Brillouin zone. It is
intrinsic and non-degenerate at room temperature with a carrier density $N$
around 10$^{22}$ m$^{-3}$. Its crystal structure consists of parallel
stacked helices of 3 Te atom per turn (Fig.\ 1) and belongs to the crystal
class \textbf{32. \FRAME{ftbpFU}{2.565in}{1.6795in}{0pt}{\Qcb{Trigonal
tellurium crystal structure. Left: view on the x-z plane of a single helix.
Right: view on the x-y plane. a = 445 pm, c = 593 pm, d = 238 pm}}{\Qlb{Fig.
1}}{Rikken-eMChA-Te-Fig1.png}{\special{language "Scientific Word";type
"GRAPHIC";maintain-aspect-ratio TRUE;display "USEDEF";valid_file "F";width
2.565in;height 1.6795in;depth 0pt;original-width 4.2851in;original-height
2.7968in;cropleft "0";croptop "1";cropright "1";cropbottom "0";filename 'Te
crystal structure new 2.png';file-properties "XNPEU";}}}The chirality of
t-Te and the absence of time reversal symmetry at the H point allow for the
existence of linear \textbf{k-}terms in the bandstructure (\textbf{k = }%
wavevector) \cite{Doi JPSJ}. Using this bandstructure, several chiroptical
properties (see \cite{Calculs} and references therein) and magneto-optical
properties (see \cite{MO props}\ and references therein) of t-Te have been
calculated, in good agreement with experiment. The existence of a strong
eMChA in t-Te is plausible because of its helical crystal structure, and it
is further supported by the recent observation of current-induced shifts in
the $^{125}$Te NMR frequency in t-Te \cite{Te NMR}. This observation
confirms the existence of strong inverse eMChA in in this material. This
effect corresponds to a current induced longitudinal magnetization present
in all chiral conductors \cite{emchaprl} and therefore implies the existence
of a strong eMChA in t-Te. Theoretical work has addressed several
contributions to the inverse eMChA in t-Te (see \cite{KME} and references
therein) but cannot be straightforwardly adapted to make quantitative
predictions for the direct eMChA effect.\newline
Single crystals of zone refined t-Te (purity better than 99,9999\%) were
commercially obtained \cite{alfa te}. They can be easily cleaved along the
c-axis, and etching the cleavage surface with hot concentrated sulfuric acid
reveals characteristic etch pits from which the crystal axis orientation and
the crystal handedness can be straightforwardly deduced \cite{etch pits Te}.
Typical sample sizes are cross-sections of 0,5 x 0,5 mm$^{2}$ and lengths of
4 mm. Gold contacts for colinear 4-terminal resistance measurements were
deposited by sputtering and a 200 Hz AC current $I^{\omega }$ was injected
with a low-distorsion current source. The generated voltages $V^{\omega }$
and its second harmonic $V^{2\omega }$ were measured by phase-sensitive
detection.\newline
The above expression for $R^{D/L}$ is a simplification, only strictly valid
in cubic crystals or isotropic media. In crystals of lower symmetry, the
correct description of eMChA requires a fourth rank tensor $\mathbf{\gamma }$%
:%
\begin{equation}
E_{i}^{2\omega }=\gamma _{ijkl}^{D/L}J_{j}^{\omega }J_{k}^{\omega }B_{l}
\label{tensor}
\end{equation}%
where $\gamma _{ijkl}^{D}=-\gamma _{ijkl}^{L}$ and $J$ is the current
density. The crystal symmetry imposes further restrictions on the tensor
components (see e.g. \cite{Birss}). \FRAME{ftbpFU}{2.29in}{3.0666in}{0pt}{%
\Qcb{Current and magnetic field dependence of the resistance (triangles,
defined as $R=V^{\protect\omega }/I^{\protect\omega }$ ) and eMChA (balls,
for definition see text) of a x-oriented left-handed Te crystal at 300 K and
with a cross section of 5 10$^{-7}$ m$^{2}$. Lines are linear fits.}}{}{b&i
result-v3.png}{\special{language "Scientific Word";type
"GRAPHIC";maintain-aspect-ratio TRUE;display "USEDEF";valid_file "F";width
2.29in;height 3.0666in;depth 0pt;original-width 6.1869in;original-height
8.2953in;cropleft "0";croptop "1";cropright "1";cropbottom "0";filename 'B&I
result-V3.png';file-properties "XNPEU";}}In the geometry used in our
experiments, $i=j=k$ and the cases for $i=x$ and $i=z$ have been addressed
by using crystals of different cuts. It can be easily shown that eMChA,
defined as $(R(B,I)-R(B,-I))/(R(B,I)+R(B,-I))\equiv \Delta R/R$ equals $%
4V^{2\omega }/V^{\omega }$ (for details see \cite{Pop}-\cite{Aoki}). By
taking the difference between the results for $+B$ and $-B$, all
nonlinearities that are even in magnetic field are eliminated. The
dependence of $\Delta R/R$ on current and magnetic field for an x-oriented
left-handed crystal (space group D$_{3}^{6}$) is shown in Figure 2,
confirming the strictly linear dependence of eMChA on these quantities. By
making different crystal cuts and by rotating the crystals, different tensor
components can be measured, as illustrated in Fig. 3 and 4. \FRAME{fhFU}{%
2.4881in}{1.7988in}{0pt}{\Qcb{eMChA of a x-oriented left handed crystal at
room temperature, as a function of the angle between $\mathbf{B}$ and the
crystal x-axis. At $\protect\theta =0%
{{}^\circ}%
$, $\Delta R/R\propto $ $\protect\gamma _{xxxx}$, at $\protect\theta =\pm 90%
{{}^\circ}%
$, $\Delta R/R\propto $ $\protect\gamma _{xxxy}$. }}{\Qlb{Fig. 3}}{%
Rikken-eMChA-Te-Fig3.png}{\special{language "Scientific Word";type
"GRAPHIC";display "USEDEF";valid_file "F";width 2.4881in;height
1.7988in;depth 0pt;original-width 5.7519in;original-height 4.5852in;cropleft
"0";croptop "1";cropright "1";cropbottom "0";filename 'MPI-6N-L result
2.png';file-properties "XNPEU";}}The enantioselectivity of the eMChA is
illustrated in Figure 4, which shows eMChA results for a left-handed and a
right handed crystal (space group D$_{3}^{4}$) in the same orientation, with
opposite results.\ \FRAME{fhFU}{2.501in}{1.7867in}{0pt}{\Qcb{eMChA of a
left-handed ($\blacktriangle $) and right-handed ($\blacktriangledown $ )
z-oriented Te crystal at room temperature, as a function of the angle
between $\mathbf{B}$ and the crystal z-axis. At $\protect\theta =0%
{{}^\circ}%
$, $\Delta R/R\propto $ $\protect\gamma _{zzzz}$, at $\protect\theta =\pm 90%
{{}^\circ}%
$,$\Delta R/R\propto $ $\protect\gamma _{zzzx}.$}}{\Qlb{Fig. 4}}{%
Rikken-eMChA-Te-Fig4.png}{\special{language "Scientific Word";type
"GRAPHIC";display "USEDEF";valid_file "F";width 2.501in;height
1.7867in;depth 0pt;original-width 5.8781in;original-height 4.5835in;cropleft
"0";croptop "1";cropright "1";cropbottom "0";filename 'D&L result
2.png';file-properties "XNPEU";}}From Figs. 3 and 4 we deduce $3\gamma
_{xxxx}\approx \gamma _{xxxy}$ , $12\gamma _{zzzx}\approx \gamma _{xxxy}$ \
and $\gamma _{zzzz}\ll \gamma _{zzzx}$. The latter result does not mean that 
$\gamma _{zzzz}$ $=0$, as there is no symmetry argument that imposes that.
The uncertainty in the exact geometrical shape of the sample, in combination
with the small value of $\gamma _{zzzz}$ and the much larger values of the
other tensor elements, make it difficult to determine a significant value
for this quantity. Fig. 5 shows the temperature dependence of eMChA in the
intrinsic regime around room temperature. \FRAME{fhFU}{2.5867in}{1.7262in}{%
0pt}{\Qcb{Temperature dependence of eMChA of an x-oriented left handed
crystal. For dotted line, see text.}}{\Qlb{Fig. 5}}{Rikken-eMChA-Te-Fig5.png%
}{\special{language "Scientific Word";type "GRAPHIC";display
"USEDEF";valid_file "F";width 2.5867in;height 1.7262in;depth
0pt;original-width 5.7372in;original-height 4.2333in;cropleft "0";croptop
"1";cropright "1";cropbottom "0";filename 'Temp dep V21.png';file-properties
"XNPEU";}}Table 1 gives a summary of the values reported so far for eMChA,
illustrating that t-Te shows by far the strongest effect.\smallskip \newline
\smallskip 
\begin{tabular}{|c|c|c|c|}
\hline
\textbf{Material \ } & $\gamma $ $[m^{2}/T\cdot A]$ & \textbf{Ref.} & 
\textbf{Remark} \\ \hline
t-Te $xxxy$ & 10$^{-8}$ & this work & RT \\ \hline
TTF-ClO$_{4}$ & 10$^{-10}$ & \cite{Pop} & RT \\ \hline
CrNb$_{3}$S$_{6}$ & 10$^{-12}$ & \cite{Aoki} & magnetic, LT \\ \hline
MnSi & 2 10$^{-13}$ & \cite{Yokouchi} & magnetic, LT \\ \hline
SWCNT & 10$^{-14}$ & \cite{eMChA CNT} & LT \\ \hline
\end{tabular}%
\smallskip \newline
\textit{Table 1 Summary of published eMChA results (RT room temperature, LT
low temperature)\newline
}\newline
We propose as explanation for the observed eMChA the \textbf{k}-linear terms
in the t-Te bandstructure around the H-point. Nakoa \textit{et al} \cite%
{Nakao JPSJ} have shown theoretically and experimentally that these terms
lead to the lifting of the energy degeneracy between valence band states
with \textbf{k} and \textbf{-k} in the presence of a magnetic field
perpendicular to the z-axis, whereas as no degeneracy lifting occurs for a
magnetic field parallel to the z-axis. Although these authors did not
comment on this aspect, such an energy-splitting has to be enantioselective
in order to be symmetry allowed. We can therefore heuristically simplify
their findings to the existence of a MChA energy term around the t-Te
valence band maximum of the form $\Delta \epsilon _{v}=\chi ^{D/L}\mathbf{k}%
\cdot \mathbf{B}_{\perp }$ with $\chi ^{D}=-\chi ^{L}$ and $\left\vert \chi
\right\vert \approx 1.5\cdot 10^{-30}\ Jm/T$. A similar behavior was later
found for the t-Te conduction band \cite{Blinowski}. If we neglect the
regular non-parabolicity of the valence band, its energy dispersion relation
around the H-point is $\epsilon (\mathbf{k})=\hbar ^{2}\mathbf{k}%
^{2}/2m^{\ast }+\chi ^{D/L}\mathbf{k}\cdot \mathbf{B}_{\perp }$. In the
constant relaxation time approximation, the Boltzmann equation gives for the
electrical conductivity $\sigma $ \cite{Seeger}:%
\begin{equation}
\sigma _{ii}=\frac{q^{2}\tau }{4\pi ^{3}\hbar ^{2}}\int \left( \frac{%
\partial \epsilon (\mathbf{k})}{\partial k_{i}}\right) ^{2}\frac{\partial f}{%
\partial \epsilon }d^{3}\mathbf{k}  \label{Conductivity}
\end{equation}%
where $f$ is the distribution function, $q$ the charge, $\tau $ the
scattering time, $m^{\ast }$ the effective mass and for Maxwell-Boltzmann
statistics, $\partial f/\partial \epsilon =-f/k_{B}T$. Applying the magnetic
field in the x-direction and neglecting terms in $B^{2}$ this results in%
\begin{eqnarray}
\sigma _{xx} &=&\frac{Nq^{2}\tau }{m^{\ast }}-\frac{q^{2}\tau \chi ^{D/L}B}{%
4\pi ^{3}m^{\ast }k_{B}T}\int k_{x}fd^{3}\mathbf{k} \\
&=&\sigma _{0}\left( 1-\gamma _{xxxx}^{D/L}J_{x}B\right) 
\end{eqnarray}%
where%
\begin{equation}
\gamma _{xxxx}^{D/L}=m^{\ast }\chi ^{D/L}/Nq\hbar k_{B}T  \label{gamma}
\end{equation}%
Note that at room temperature an analogous contribution to eMChA will come
from the conduction band which, depending on the relative signs of $\chi $
for the two bands, may have the same or opposite sign. This electron
contribution will disappear at low temperature when t-Te becomes extrinsic
p-type. This simple model leads to several predictions:\newline
1) as a general rule, metals, with their much larger carrier density, will
show smaller eMChA than semiconductors and semi-metals, in agreement with
the table above.\newline
2) for t-Te at 300 K, $\Delta R/R(\mathbf{B}//\mathbf{J}//\mathbf{x})=$\ $%
\gamma _{xxxx}JB$ with $\gamma _{xxxx}\approx 6.10^{-10}$ $m^{2}A^{-1}T^{-1}$
whereas our experimental result (Fig. 3) gives $\gamma _{xxxx}\approx
2.10^{-9}$ $m^{2}A^{-1}T^{-1}$, a reasonable agreement in view of the\
simple model. \newline
3) $\gamma _{iiiz}=0$ in t-Te because of the absence of the degeneracy
lifting for $\mathbf{B}\parallel \mathbf{z}$, in agreement with our
observations. Other mechanisms for eMChA \cite{emchaprl}, not included in
our model, may still lead to a small non-zero value. One may obtain an
estimate for $\gamma _{zzzz}$ from the free-electron-on-a-helix model \cite%
{Free electron helix}. This can only give an upper limit, as there is
significant coupling between adjacent Te helices and charge carriers are
therefore not confined to one helix. For the same parameters as in Fig. 4,
this model predicts $\Delta R/R\approx 10^{-9}$, consistent with the
experimental result.\newline
4) Taking into account the temperature dependence of the carrier
concentration \cite{Seeger}, Eq. \ref{gamma} implies the temperature
dependence of $\gamma _{xxxx}$ to be $T^{-5/2}\exp (E_{g}/2k_{B}T).$ Fig. 5
shows that this is a reasonably good description.\newline
Although t-Te is topologically trivial, the existence of \textbf{k}-linear
terms in its bandstructure and its strong spin-orbit interaction suggest a
link to topological insulators and Weyl semi-metals. Indeed, strong eMChA
was also claimed for topologically non-trivial systems like Weyl semi-metals 
\cite{Morimoto} and non-centrosymmetric Rashba superconductors \cite%
{Wakatsuki}, although these systems are not chiral in the strict sense of
the term. In line with this suggestion, t-Te is predicted to become a strong
topological insulator under pressure \cite{Agapito} and ARPES measurements
have revealed Weyl nodes at ambient pressure \cite{Nakayama}. \newline
It will be clear that a more sophisticated theoretical approach, including
the effect of the Lorentz force, plus the full details of the band structure
are necessary to obtain full understanding of eMChA in t-Te and to arrive at
an accurate description of all its tensor elements. Our model allows to
identify the major contributing factors and  allows to identify other
materials showing strong eMChA through straightforward band structure
calculations, thereby opening a venue for realistic applications of this
effect. Obvious candidates are trigonal selenium and its binary alloys with
tellurium, and cinnabar ($\alpha -$HgS) that have the same crystal
structure, albeit slightly different band structures. At different band
structure extrema, the MChA energy term, present around the H point in t-Te,
may be zero, which could lead to a much lower eMChA.\newline
In summary, we have experimentally observed strong eMChA in t-Te, we have
demonstrated the tensorial character of this effect and we have identified a
novel intrinsic bandstructure mechanism that agrees with these results.

This work was supported by the French National Agency for Research
(ChiraMolCo, ANR 15-CE29-0006-02).\

\end{document}